\documentclass[aps,prc,showpacs,preprintnumbers]{revtex4}%
\usepackage{amssymb}
\usepackage{amsmath}
\usepackage{amsfonts}
\usepackage{graphicx}
\usepackage{bm}%
\providecommand{\U}[1]{\protect\rule{.1in}{.1in}}
\providecommand{\U}[1]{\protect\rule{.1in}{.1in}}
\begin{document}
\title{    Probing two-photon decay widths of mesons at energies available
    at the CERN Large Hadron Collider (LHC)}
\author{C.A. Bertulani}
\address{Department of Physics, Texas A\&M University-Commerce, TX
75429}

\begin{abstract}
Meson production cross sections in ultra-peripheral
relativistic heavy ion collisions at the LHC are revisited. The relevance of
meson models and of exotic QCD states is discussed. This study includes  states that have not been considered
before in the literature.

\end{abstract}
\date{\today }

 \keywords{}\maketitle

In quantum chromodynamics (QCD), meson spectroscopy is an exciting
possibility to search for states which cannot be explained within
the $q\overline{q}$ model. A few possibilities are multiquark states
such as molecules $(q\overline{q})(q\overline{q})$, hybrid mesons
$(q\overline{q}g)$ and glueballs $(gg)$. Despite the intense the
considerable experimental effort such states have yet to be
established \cite{Yao06}. The masses predicted for these states are
typically in the range of 1.5 to 2.5 GeV. Therefore, a crucial part
of the search for these ``abnormal" states is to establish the
spectrum of ``ordinary" $q\overline{q}$-mesons in the 1.5-2.5 GeV
region. The ``abnormal" states may only stand out clearly when the
ordinary $q\overline{q}$ states are a well understood and classified
background \cite{Bar85}. This has been the subject of intense
investigation for a long time (see, e.g.
\cite{JJ76,Jaf77,Jaf78,Bar85,Bar92,Bar92b,Bar92c,Bar92d,Bar97}).

Two-photon physics can contribute to the search for non-$q\overline{q}$
resonances both by establishing the spectrum of $q\overline{q}$ levels and by
identifying states with anomalous $\gamma\gamma$ couplings. $\gamma\gamma$
collisions are useful for determining the spectrum of $\gamma\gamma$ states
because $q\overline{q}$ mesons have a characteristic pattern of $\gamma\gamma$
couplings, for example the relative flavor factor within a flavor-SU(3)
multiplet of $q\overline{q}$ states:
\begin{equation}
\Gamma_{\gamma\gamma}(f) \ :\ \Gamma_{\gamma\gamma}(a) \ :\ \Gamma
_{\gamma\gamma}(f^{\prime}) = 25 \ : \ 9 \ : \ 2,
\end{equation}
which, except for minor relativistic corrections, reproduce the experimental
data quite well. If there were any doubt about the identity of one of the
light $^{3}P_{2}$ states $f_{2}(1270)$, $a_{2}(1320)$ and $f_{2}^{\prime
}(1525)$, for example, their relative, $\gamma\gamma$ widths would be a strong
argument that these are all $q\overline{q}$ states if any one had been
established as $q\overline{q}$. The $\Gamma_{\gamma\gamma}$ partial widths of
resonances would be even more useful in the identification of $q\overline{q}$
states if the absolute scale of these widths can be reliably calculated. But
this proves to be rather sensitive to the assumptions made in calculation, as
has been noted in several theoretical investigations.

There is a true motivation to observe if predictions of ``abnormal" states
within the quark model is verified experimentally and complies with the
$\Gamma_{\gamma\gamma}$ partial widths calculated for ``normal states". This
fact can only be assessed experimentally. Photon-photon (or \textquotedblleft
two-photon\textquotedblright) processes have long been studied at $e^{+}e^{-}$
colliders \cite{Bud74}. They are an excellent tool for investigating many
aspects of meson spectroscopy, and allow intensity tests of QED. At the
highest energy colliders, reactions such as $\gamma\gamma\rightarrow X$ may be
used to probe the quark content and spin structure of meson resonances.
Production of meson or baryon pairs can also probe the internal structure of hadrons.

A wonderful new possibility for similar experimental studies is the Large
Hadron Collider (LHC) at CERN. At the LHC, photon-proton collisions occur at
center of mass energies an order of magnitude higher than were available at
previously existing accelerators, and photon-heavy ion collisions reach 30
times the energies available at fixed target accelerators. The Lorentz gamma
factor $\gamma=(1-v^{2}/c^{4})^{-1/2}$ in the laboratory frame is 7000 for
p-p, 3000 for Pb-Pb collisions. Due to the ions large charges (e.g., $Z=82$)
and their short-interaction time ($\Delta t \simeq20 \gamma$ MeV), the
interacting electromagnetic fields generated by these ions are typically much
stronger ($\propto Z^{2}$) than the Schwinger critical field
\cite{Sch49,Bau08} $E_{Sch} = m^{2}/he = 1.3 \times10^{16}$ V cm. Light
particles (e.g., $e^{+}e^{-}$-pairs are produced copiously by such fields
\cite{BB88}. The electromagnetic fields of heavy-ions are of sufficient
intensity to allow the study of multi-photon reactions (for an early review,
see ref. \cite{BB88}). At the LHC, electroweak processes such as $\gamma
\gamma\rightarrow W^{+}W^{-}$ may also be studied. Even the production of the
Higgs boson is not negligible with such a mechanism \cite{Bau88,Pap89,Gra89}.
The physics of ultra-peripheral heavy ion collisions has become object of
intensive studies in recent years
\cite{KGS97,BHT02,BKN05,BHT07,Ba08,Ba08b,Nyst08}.

In the equivalent photon approximation, for ion beams with squared center of
mass energy $s$, one can write the cross section for photon-photon fusion to a
neutral state $X$ in the form \cite{BB88}
\begin{equation}
\sigma_{X}=\int dx_{1}dx_{2} N_{\gamma}\left(  x_{1}\right)  N_{\gamma}\left(
x_{2}\right)  \sigma^{X}_{\gamma\gamma}\left(  x_{1}x_{2}s\right)  \;,
\label{eq:two-photon}%
\end{equation}
where $N_{\gamma}(x)$ is the distribution function (equivalent photon numbers)
for finding a quantum $\gamma$ with energy fraction $x$ and $\sigma
^{X}_{\gamma\gamma}\left(  x_{1}x_{2}s\right)  $ is the two-photon cross
section. It is given by \cite{Low60}
\begin{equation}
\sigma^{X}_{\gamma\gamma}\left(  x_{1}x_{2}s\right)  =8\pi^{2}(2J+1){\frac
{\Gamma_{m_{X}\rightarrow\gamma\gamma}}{m_{X}}}\ \delta\left(  x_{1}x_{2}s
-m^{2}_{X}\right)  \label{Low}%
\end{equation}
where $J$, $m_{X}$, and $\Gamma_{m_{X}\rightarrow\gamma\gamma}$ are the spin,
mass and the two-photon partial decay width. The delta-function enforces
energy conservation.

The $\gamma$-nucleus vertex is given by $ZeF(t)$, where $F(t)$ is the elastic
nuclear form factor and $t$ is the invariant four-momentum exchanged. Then the
distribution function $N_{\gamma}(x)$ for a fast moving nucleus of charge $Z$
is given by \cite{BHT02}
\begin{equation}
N_{\gamma}(x)={\frac{Z^{2}\alpha}{\pi x}} \int_{0}^{\infty}dk^{2}
\ k^{2}\ {\frac{|F(x^{2}M^{2}+k^{2})|^{2} }{(x^{2}M^{2}+k^{2})^{2}}}.
\end{equation}
If one approximates the form factor by a gaussian, $F(k^{2})=\exp
(-k^{2}/2K^{2})$, one sees that the form factor imposes a cutoff $xM/K \sim
xMR \sim< 1$, where $R \sim1/K$ is the nuclear radius. Hence, an state of
invariant mass $m_{X}$ can be produced as long as $m^{2}_{X} = x_{1}x_{2}s <
s/M^{2}R^{2} = (2\gamma/R)^{2}$. This is essentially the condition for
coherence: the photon wavelength must be larger than the Lorentz-contracted
nuclear radius. Coherence leads to the factor of $Z^{2}$ in $N_{\gamma}$
(yielding a factor $Z^{4}$ in the cross section), rendering electromagnetic
interactions of high-Z ions an effective tool for the production of heavy
neutral particles.

The above formulation does not account for the effects of inelastic nuclear
scattering, which is more easily taken into account in the impact parameter
semiclassical method. The majority of the inelastic events occur at small
values of the impact parameter $b$. The elastic nature of the interaction is
maintained only in those collisions in which the two nuclei pass far from each
other. The impact parameter representation of $N_{\gamma}(x)$ given by
\cite{BB88}
\begin{equation}
N_{\gamma}(\omega,b)=\frac{Z\alpha\omega}{\pi^{2}\gamma^{2}}\left[  K_{1}%
^{2}(y)+\frac{1}{\gamma^{2}}K_{0}^{2}(y)\right]  . \label{eq:ww1}%
\end{equation}
where $\omega=x E$ is the energy of the scattered quanta ($E$ is the ion
energy), $y=\omega b/\gamma$, $\alpha=1/137$, and $K_{0}$ and $K_{1}$ are
modified Bessel functions. The first term, $K_{1}^{2}(y)$, gives the flux of
photons transversely polarized to the ion direction and the second is the flux
for longitudinally polarized photons. The photon flux is exponentially
suppressed when $\omega>\gamma/b$. These photons are almost real, with
virtuality $-q^{2}<1/R^{2}$. The usable photon flux depends on the geometry.
Most peripheral reactions lead to final states with a handful of particles.
These final states will be overwhelmed by any hadronic interactions between
the fast moving ion and the target. Thus, the useful photon flux is that for
which the ions do not overlap, i.e. when the impact parameter $b=|\mathbf{b}%
_{1}-\mathbf{b}_{2}|$ is greater than twice the nuclear radius ($2R$) .
Usually, we can take $R=1.2A^{1/3}$ fm, where $A$ is the atomic number. The
photons can interact with a target nucleus in a one-photon process, (when
$b_{1}<R$) or with its electromagnetic field in a two-photon process when
$b_{1}>R$ and $b_{2}>R$.

For two-photon exchange processes, the equivalent photon numbers, in equation
\ref{eq:two-photon} must account for the electric field orientation of the
photon fluxes with respect to each ion \cite{BF90}, obeying the ion
non-overlap criteria $b_{1},\ b_{2}>R_{A}$. Owing to symmetry properties,
$J^{\pi}=0^{+}$ (scalar) particles originate from configurations such that
$E_{1}\parallel E_{2}$, whereas $0^{-}$ (pseudo-scalar) particles originate
from $E_{1}\perp E_{2}$ \cite{BF90,Vid93}.
\begin{equation}
\sigma_{X}=\int d\omega_{1}d\omega_{2}\int_{b_{1}>R}\int_{b_{2}>R}d^{2}%
b_{1}d^{2}b_{2} N_{\gamma}(\omega_{1},b_{1}) N_{\gamma}(\omega_{2},b_{2})
\sigma_{X}^{\gamma\gamma}\left(  \omega_{1},\omega_{2}\right)  \;,
\label{mesprod}%
\end{equation}
with the condition that $\left\vert \mathbf{b}_{1}-\mathbf{b}_{2}\right\vert
\geq2R$.

We use the formalism described above together with the $\Gamma_{\gamma\gamma}$
widths either taken from experiment, or from theory, to generate tables 1-4.
For many cases, there is no major difference between our values and the ones
obtained in previous calculations for the same mesons
\cite{JJ76,Jaf77,Jaf78,Bar85,Bar92,Bar92b,Bar92c,Bar92d,Bar97}). However, many
of our values are predictive and have not been considered before. In these
tables the properties of some $q\overline{q}$ states are given, and their
production cross sections are predicted for Pb-Pb collisions at the LHC. Mass
values and known widths are taken from theory and from experiment when
possible. Ion luminosities of $10\times26$ cm$^{-2}$ s$^{-1}$ for Pb-Pb
collisions at LHC lead up to million of events (e.g. charmonium states) per
second for the largest cross sections \cite{BHT02}. The two-photon width is a
probe of the charge of its constituents, so the magnitude of the two-photon
coupling can serve to distinguish quark dominated resonances from
glue-dominated resonances (``glueballs''). The absence of meson production via
$\gamma\gamma$ fusion is a signal of great interest for glueball search. In
ion-ion collisions, a glueball can only be produced via the annihilation of a
$q\overline{q}$ pair into gluons pairs, whereas a normal $q\overline{q}$ meson
can be produced directly. Due to the copious production of such states in
peripheral collisions at the LHC, such studies will be viable depending on the
detection setup \cite{Ba08b}. It requires measuring events characterized by
relatively small multiplicities and a small background (especially when
compared with the central collisions).

In conclusion, in this brief report, the cross sections for
meson production in ultra-peripheral collisions of heavy-ions at LHC have been reviewed and additional cases have also been considered. Due to
the very strong electromagnetic fields of short duration, new possibilities
for interesting physics arise at the LHC. The method of equivalent photons is a well
established tool to describe this kind of reactions. But, unlike electrons
and positrons, heavy ions and protons are particles with internal structure.
Thus, effects arising from this structure have to be controlled and minor
uncertainties coming from the exclusion of central collisions and triggering
must be eliminated, e.g., by using a luminosity monitor from $\mu$-- or $e$--pairs
\cite{Ba08b}. Ultra-peripheral heavy ion collisions at the LHC  is an excellent tool to produce and study 
``abnormal" $q\overline{q}$ states and/or glueballs. This encompasses
invariant masses up to 10 GeV. The production via photon-photon fusion
complements the production from single timelike photons in $e^{+}$--$e^{-}$
collider and also in hadronic collisions via other partonic processes. The
extent to which the measured production cross sections agrees with results
presented here may serve as a measure of the status of several ``anomalous"
$q\overline{q}$ and glueballs candidacies. One must reiterate that it is in general
a good strategy to study $\gamma\gamma$ meson decays through the inverse
process, i.e $\gamma\gamma$-production in heavy-ion colliders. The rate of
production of these mesons can probe the relative couplings of different decay
modes which are usually quite distinct for hybrid versus quarkonium assignments.

\medskip The author is grateful to Ted Barnes for beneficial discussions.  This work was partially supported by the U.S. DOE grants
DE-FG02-08ER41533 and DE-FC02-07ER41457 (UNEDF, SciDAC-2).

\section{Tables}

\begin{center}%
\begin{tabular}
[c]{llllll}\hline\hline
Mesons $a$ and $f$ & $J^{PC}$ & $\Gamma_{\gamma\gamma}^{th}$ [keV] &
$\Gamma_{\gamma\gamma}^{\exp}$ [keV] & Obs. & $\sigma_{\gamma\gamma}^{X}%
$\\\hline
a$_{0}^{K\overline{K}}\left(  980\right)  $ & (0$^{++}$) & 0.6 & $0.30\pm0.10$
& $\longrightarrow K\overline{K}\longrightarrow\gamma\overline{\gamma}$ & 3.1
mb\\
a$_{0}^{q\overline{q}}\left(  980\right)  $ &  & 1.5 &  & hypothetical, NR
q-model & 8.6 mb\\
a$_{0}^{q\overline{q}}\left(  980\right)  $ &  & 1.0 &  & hypothetical, R
q-model & 5.5 mb\\
f$_{0}^{K\overline{K}}\left(  980\right)  $ & (0$^{++}$) & 0.6 &
$0.29_{-0.09}^{+0.07}$ & $\longrightarrow K\overline{K}\longrightarrow
\gamma\overline{\gamma}$ & 3.1 mb\\
f$_{0}^{q\overline{q}}\left(  980\right)  $ &  & 4.5 &  & hypothetical, NR
q-model & 25.8 mb\\
f$_{0}^{q\overline{q}}\left(  980\right)  $ &  & 2.5 &  & hypothetical, R
q-model & 14.3 mb\\
f$_{0}\left(  1200\right)  $ &  & 3.25 - 6.46 & unknown & for $m_{q}=0.33$ to
0.22 GeV & 9.6 - 21 mb\\
f$_{2}(1274)$ & (2$^{++}$) & 1.75 - 4.04 & $2.6\pm0.24$ & $\Gamma
_{\gamma\gamma}$ $\left(  \text{f}_{0}\right)  /\Gamma_{\gamma\gamma}$
$\left(  \text{f}_{2}\right)  =$ 1.86 - 1.60 & 21 - 49 mb\\
f$_{2}^{\lambda=2}\left(  1274\right)  $ &  & 1.71 - 3.93 &  & $\left(
\lambda=0\right)  /\left(  \lambda=2\right)  =$ 0.022 - 0.029 & 20 - 44 mb\\
f$_{2}^{\lambda=0}\left(  1274\right)  $ &  & 0.04 - 0.11 &  &  & 0.09 - 0.23
mb\\
f$_{0}\left(  1800\right)  $ &  & 2.16 - 2.52 & unknown & 2$^{3}$P$_{0}$
radial excitation & 2.5 - 3.1 mb\\
f$_{2}\left(  1800\right)  $ $\left(  \lambda=2\right)  $ &  & 1.53 - 2.44 &
& 2$^{3}$P$_{2}$ radial excitation & 1.7 - 2.9 mb\\
f$_{2}\left(  1800\right)  $ $\left(  \lambda=0\right)  $ &  & 0.08 - 0.16 &
& " & 0.08 - 14 mb\\
f$_{2}\left(  1525\right)  $ & (2$^{++}$) & 0.17 & $0.081\pm0.009$ &
$s\overline{s},$ $m_{s}=0.55$ GeV fixed & 0.86 mb\\
f'$_{2}\left(  1525\right)  \left(  \lambda=0\right)  $ &  & 0.065 &  & " &
0.21 mb\\
f'$_{2}\left(  1525\right)  \left(  \lambda=2\right)  $ &  & $0.9\times
10^{-3}$ &  & " & 0.42 $\mu$b\\
f$_{4}\left(  2050\right)  $ &  & 0.36 - 1.76 & unknown & $^{3}$F$_{4}$ & 0.03
- 0.14 mb\\
f$_{4}\left(  2050\right)  \left(  \lambda=2\right)  $ &  & 0.33 - 1.56 &  &
" & 0.02 - 0.13 mb\\
f$_{4}\left(  2050\right)  \left(  \lambda=0\right)  $ &  & 0.03 - 0.20 &  &
" & 2 - 12 $\mu$b\\
f$_{3}\left(  2050\right)  $ &  & 0.50 - 2.49 & unknown & $^{3}$F$_{3}$ & 0.03
- 0.13 mb\\
f$_{2}\left(  2050\right)  $ &  & 2.48 - 11.11 & unknown & $^{3}$F$_{2}$ &
0.12 - 0.53 mb\\
f$_{2}\left(  2050\right)  \left(  \lambda=2\right)  $ &  & 1.85 - 8.49 &  &
" & 0.09 - 0.46 mb\\
f$_{2}\left(  2050\right)  \left(  \lambda=0\right)  $ &  & 0.63 - 2.62 &  &
" & 0.01 - 0.07 mb\\
f$_{0}^{K^{\ast}K^{\ast}}\left(  \simeq1750\right)  $ &  & $\simeq0.05-0.1$ &
unknown & vector-vector molecule & 0.19 mb\\\hline\hline
\label{t1} &  &  &  &  &
\end{tabular}

\end{center}

Table 1. $\gamma-\gamma$ widths for mesons $a$ and $f$ calculated with the
models described in refs.
\cite{JJ76,Jaf77,Jaf78,Bar85,Bar92,Bar92b,Bar92c,Bar92d}. Masses above 1250
MeV are assumed within parenthesis. Experimental values of the $\gamma-\gamma$
widths are extracted form the Particle Data Properties Web site.

\begin{center}%
\begin{tabular}
[c]{llllll}\hline\hline
Mesons $\eta,$ $\chi$ and $h$ $(c\overline{c})$ & $J^{PC}$ & $\Gamma
_{\gamma\gamma}^{th}$ [keV] & $\Gamma_{\gamma\gamma}^{\exp}$ [keV] & Obs. &
$\sigma_{\gamma\gamma}^{X}$\\\hline
$\eta_{c}$ & (0$^{-+}$) & 3.4 - 4.8 & $6.7_{-0.8}^{+0.9}$ & $m_{c}=1.4-1.6$
GeV & 0.26 - 0.34 mb\\
$\eta_{c}(3790)$ &  & 1.85 - 8.49 & $1.3\pm0.6$ & $m_{c}=1.4$ GeV & 0.06 - 0.1
mb\\
$\eta_{c}^{\prime}(3790)$ &  & 3.7 & unknown & $m_{c}=1.4$ GeV & 0.11 mb\\
$\eta_{c}(4060)$ &  & 3.3 & unknown &  & 0.09 mb\\
$\eta_{c2}^{1D}(3840)$ &  & $20.\times10^{-3}$ & unknown &  & 0.15 $\mu$b\\
$\eta_{c2}^{2D}(4210)$ &  & $35.\times10^{-3}$ & unknown &  & 0.14 $\mu$b\\
$\eta_{c4}^{1G}(4350)$ &  & $0.92\times10^{-3}$ & unknown &  & 0.08 $\mu$b\\
$\chi_{2}$ & (2$^{++}$) & 0.56 & $0.258\pm0.019$ & $\left(  \lambda=2\right)
/\left(  \lambda=0\right)  =0.005$ & 82 $\mu$b\\
$\chi_{0}$ & (0$^{++}$) & 1.56 & $0.276\pm0.033$ & $\Gamma_{\gamma\gamma}$
$\left(  \chi_{0}\right)  /\Gamma_{\gamma\gamma}$ $\left(  \chi_{2}\right)
=2.79$ & 0.05 mb\\
$\chi_{2}^{\prime}$ & (2$^{++}$) & 0.64 & unknown &  & 0.09 mb\\
h$_{c2}(3840)$ &  & $20\times10^{-3}$ & unknown & $^{1}$D$_{2}$ & 82 $\mu$b\\
$\chi_{2}\left(  4100\right)  $ &  & $30\times10^{-3}$ & unknown & $^{3}%
$F$_{2}$ & 0.11 $\mu$b\\\hline\hline
\label{t2} &  &  &  &  &
\end{tabular}

\end{center}

Table 2. $\gamma-\gamma$ widths for $c\bar{c}$-mesons $\eta$, $\chi$ and $h$
calculated with the models described in refs.
\cite{JJ76,Jaf77,Jaf78,Bar85,Bar92,Bar92b,Bar92c,Bar92d}. Masses above 1250
MeV are assumed within parenthesis. Experimental values of the $\gamma-\gamma$
widths are extracted form the Particle Data Properties Web site.

\bigskip

\bigskip

\begin{center}%
\begin{tabular}
[c]{llllll}\hline\hline
Mesons $\eta,$ $\chi$ and $h$ ($b\overline{b}$) & $J^{PC}$ & $\Gamma
_{\gamma\gamma}^{th}$ [keV] & $\Gamma_{\gamma\gamma}^{\exp}$ [keV] & Obs. &
$\sigma_{\gamma\gamma}^{X}$\\\hline
$\eta_{b}^{1S}(9400)$ &  & $0.17\times10^{-3}$ & unknown &  & 19 nb\\
$\eta_{b}^{2S}(9400)$ &  & $0.13\times10^{-3}$ & unknown &  & 16 nb\\
$\eta_{b}^{3S}(9480)$ &  & $0.11\times10^{-3}$ & unknown &  & 14 nb\\
$\eta_{b2}^{1D}(10150)$ &  & $33.\times10^{-6}$ & unknown &  & 0.4 nb\\
$\eta_{b2}^{2D}(10450)$ &  & $69.\times10^{-6}$ & unknown &  & 0.8 nb\\
$\eta_{b4}^{1G}(10150)$ &  & $59.\times10^{-6}$ & unknown &  & 0.7 nb\\
$\eta_{b}(9366)$ & (0$^{-+}$) & 0.17 & unknown &  & 0.12 $\mu$b\\
$\eta_{b}^{\prime}$ &  & 0.13 & unknown &  & 0.17 $\mu$b\\
$\eta_{b}^{\prime\prime}$ &  & 0.11 & unknown & $s\overline{s},$ $m_{s}=0.55$
GeV fixed & 0.15 $\mu$b\\
$\chi_{b2}(9913)$ & (2$^{++}$) & $3.7\times10^{-3}$ & unknown &  & 0.09 $\mu
$b\\
$\chi_{b0}(9860)$ & (0$^{++}$) & $13.\times10^{-3}$ & unknown &  & 0.08 $\mu
$b\\\hline\hline
\label{t3} &  &  &  &  &
\end{tabular}

\end{center}

Table 3. $\gamma-\gamma$ widths for $b\bar{b}$-mesons $\eta$ and $\chi$ calculated
with the models described in refs.
\cite{JJ76,Jaf77,Jaf78,Bar85,Bar92,Bar92b,Bar92c,Bar92d}. Masses above 1250
MeV are assumed within parenthesis. Experimental values of the $\gamma-\gamma$
widths are extracted form the Particle Data Properties Web site.

\bigskip

\bigskip

\begin{center}%
\begin{tabular}
[c]{llllll}\hline\hline
Mesons $\pi$ & $J^{PC}$ & $\Gamma_{\gamma\gamma}^{th}$ [keV] & $\Gamma
_{\gamma\gamma}^{\exp}$ [keV] & Obs. & $\sigma_{\gamma\gamma}^{X}$\\\hline
$\pi_{0}$ & (0$^{-+}$) & $3.4-6.4\times10^{-3}$ & $8.4\pm0.6\times10^{-3}$ &
$m_{q}=0.220-.33$ GeV & 27 - 52 m b\\
$\pi(1300)$ & (0$^{-+}$) & $0.43-0.49$ & unknown &  & 0.69 - 0.71 mb\\
$\pi(1880)$ &  & $0.74-1.0$ & unknown & 3$^{1}$S$_{0}$ & 0.8 - 1.1 mb\\
$\pi_{2}(1670)$ & (2$^{-+}$) & $0.11-0.27$ & $<0.072$ &  & 0.41 - 1.1 mb\\
$\pi_{2}^{\prime}(2130)$ & (2$^{-+}$) & $0.10-0.16$ & unknown & 2$^{1}$D$_{2}$
& 0.36 - 0.54 mb\\
$\pi_{4}(2330)$ &  & $0.21-1.6$ & unknown & $^{1}$G$_{4}$ & 0.04 - 0.31
mb\\\hline\hline
\label{t4} &  &  &  &  &
\end{tabular}

\end{center}

Table 4. $\gamma-\gamma$ widths for $\pi$-mesons calculated with the models
described in refs. \cite{JJ76,Jaf77,Jaf78,Bar85,Bar92,Bar92b,Bar92c,Bar92d}.
Masses above 1250 MeV are assumed within parenthesis. Experimental values of
the $\gamma-\gamma$ widths are extracted form the Particle Data Properties Web site.

\end{document}